\documentclass[aps,twocolumn,groupedaddress,showpacs]{revtex4}
\usepackage{graphicx}
\begin{document}
%%%%%%%%%%%%%%%%%%%%%%%%%%%%%%%%%%%%%%%%%%%%%%%%%%%%%%%%%%%%%%%%%%%%%%%%%%%%%
% You should use BibTeX and revtex.bst for references
%\bibliographystyle{apsrev}
%%%%%%%%%%%%%%%%%%%%%%%%%%%%%%%%%%%%%%%%%%%%%%%%%%%%%%%%%%%%%%%%%%%%%%%%%%%%%
% marks overfull lines with blackboxes
%\draft - no longer supported, use the 'draft' option instead
%\preprint{}
\title{Numerical study of magnetization processes in rare-earth tetraborides}
% Optional argument for running titles on pages
%\title[]{}
\author{Pavol Farka\v{s}ovsk\'y, Hana \v{C}en\v{c}arikov\'a and Slavom\'ir
Mat$\!$'a\v{s}}
%\thanks{}
\affiliation{Institute of Experimental Physics, Slovak Academy
of Sciences, Watsonova 47, 040 01 Ko\v{s}ice, Slovakia}

\date{\today}
%%%%%%%%%%%%%%%%%%%%%%%%%%%%%%%%%%%%%%%%%%%%%%%%%%%%%%%%%%%%%%%%%%%%%%%%%%%%%
%                         ABSTRACT                                          %
%%%%%%%%%%%%%%%%%%%%%%%%%%%%%%%%%%%%%%%%%%%%%%%%%%%%%%%%%%%%%%%%%%%%%%%%%%%%%
\begin{abstract}
We present a simple model for a description of  magnetization processes in 
rare-earth tetraborides. The model is based on the coexistence of two 
subsystems, and namely, the spin subsystem described by the Ising model 
and the electronic subsystem described by the Falicov-Kimball model on the 
Shastry-Sutherland lattice (SSL). Moreover, both subsystems are coupled by 
the anisotropic spin-dependent interaction of the Ising type. 
We have found, that 
the switching on the spin-dependent interaction ($J_z$) between the electron and 
spin subsystems 
and taking into account the electron hopping on the nearest 
($t$) 
and next-nearest ($t'$) lattice sites of the SSL leads to a stabilization of new 
magnetization plateaus. 
In addition, to the Ising magnetization plateau at 
$m^{sp}/m_s^{sp}=1/3$ we have found three new magnetization plateaus 
located at $m^{sp}/m_s^{sp}=1/2$, 1/5 and 1/7 of the 
saturated spin magnetization $m_s^{sp}$. The ground-states corresponding to 
magnetization plateaus have the same spin structure consisting of
parallel antiferromagnetic bands separated by ferromagnetic~stripes.
\end{abstract}
\pacs{75.10.-b,75.60.Ej,75.40.Mg}
%%%%%%%%%%%%%%%%%%%%%%%%%%%%%%%%%%%%%%%%%%%%%%%%%%%%%%%%%%%%%%%%%%%%%%%%%%%%%
%\maketitle must follow title, authors, abstract and \pacs
\maketitle
% References should be done using the \cite, \cite, and \label commands
%\section{}
%\label{}
%\subsection{}
%\subsubsection{}
%%%%%%%%%%%%%%%%%%%%%%%%%%%%%%%%%%%%%%%%%%%%%%%%%%%%%%%%%%%%%%%%%%%%%%%%%%%%%
%                        MAIN TEXT                                          %
%%%%%%%%%%%%%%%%%%%%%%%%%%%%%%%%%%%%%%%%%%%%%%%%%%%%%%%%%%%%%%%%%%%%%%%%%%%%%
\section{INTRODUCTION}
%%%%%%%%%%%%%%%%%%%%%%%%%%%%%%%%%%%%%%%%%%%%%%%%%%%%%%%%%%%%%%%%%%%%%%%%%%%
The Shastry-Sutherland lattice (SSL) was considered more than 20 years ago 
by Shastry and Sutherland~\cite{Shastry} as an interesting example of a 
frustrated quantum spin system with an exact ground state. It can be 
described as a square lattice with antiferromagnetic couplings $J$ between 
nearest neighbors and additional antiferromagnetic couplings $J'$ between 
next-nearest neighbors in every second square (see Fig.~1). 
This lattice 
attracted much attention after its experimental realization in the 
$SrCu_2(BO_3)_2$ compound~\cite{SrCu}. The observation of a fascinating sequence
of magnetization ($m/m_s=$1/2, 1/3, 1/4 and 1/8 of the saturated
magnetization $m_s$) in this material~\cite{SrCu} stimulated 
further theoretical and experimental studies of the
SSL~\cite{Sebastian,Dorier}.

	As another realization of the SSL the rare-earth tetraborid $TmB_4$ has recently
been studied in finite magnetic fields~\cite{Gabani}. Since fully polarized state 
can be reached for experimentally accessible magnetic fields, this compound 
allows exploration of its complete magnetization process. It was found that the 
magnetization diagram of $TmB_4$ consists of magnetization plateaus located 
at small fractional values of $m/m_s=$1/7, 1/8, 1/9 $\dots$ of the saturated 
magnetization, followed by the major magnetization plateau located at 
$m/m_s=1/2$. Note that, due to large total magnetic moments of the magnetic ions, this 
compound can be considered as a classical system. Moreover, because of strong 
crystal field effects, the effective spin model for $TmB_4$ has been suggested 
to be described by the spin-1/2 Shastry-Sutherland model under strong 
Ising (or easy-axis) anisotropy~\cite{Gabani}. 
From this point of view it was natural to begin a description 
of magnetization process in the $TmB_4$ material from the Ising limit 
on the SSL that can be, in the presence of a finite magnetic field $h$, expressed 
as follows
\begin{equation}
H_{JJ'}=J\sum_{\langle i,j\rangle} S^z_iS^z_j + J'\sum_{\langle\langle
 i,j\rangle\rangle} S^z_iS^z_j - h\sum_i S^z_i\ ,
\label{eq1}
\end{equation}
where $S^z_i=\pm1/2$ denotes the $z$-component of a spin-1/2 degree of freedom on 
site $i$ of a square lattice and $J$, $J'$ are the antiferromagnetic exchange 
couplings between all nearest neighbor bonds ($J$) and next-nearest neighbor 
bonds in every second square ($J'$), as indicated in Fig.~1.\\
%%%%%%%%%fig1%%%%%%%%%%%%
\vspace*{-0.5cm}
\begin{figure}[h]
\begin{center}
\includegraphics*[width=6cm]{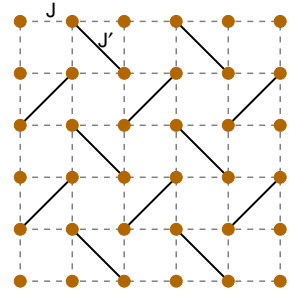}
\end{center}
\vspace*{-0.5cm}
\caption{(Color online) The Shastry-Sutherland lattice with magnetic couplings
$J$ bonds along the edges of the squares and $J'$ along the diagonals.}
\end{figure}
%%%%%%%%%%%%%%%%%%%%%%

In spite the relative simplicity of the model Hamiltonian (1), fully different 
conclusions have been obtained for the magnetization curve of this model 
within various approaches. For example, the authors of Ref.~\cite{Gabani} 
found, analyzing 
a finite system consisting of 16 spins, a single magnetization plateau at 1/2 
of the saturated magnetization in accordance with experimental data in $TmB_4$.
However, numerical simulations obtained within the Monte-Carlo and tensor 
renormalization group methods on much larger systems~\cite{Meng,Chang} did not confirm 
this conclusion. In contrast to previous results they showed that the Ising 
model on the SSL exhibits in the presence of the magnetic field the magnetization 
plateau only at 1/3 of the saturated magnetization. Thus the different 
conclusion of Ref.~\cite{Gabani} appears to be due to the usage of
inappropriate finite 
lattice sizes. 

The existence of the magnetization plateau at only 1/3 of the saturated 
magnetization and its absence at 1/2 indicates 
that it is necessary to go beyond the classical Ising limit to reach the 
correct description of the magnetization process in $TmB_4$ and other rare-earth 
tetraborides. The first such an attempt has been done by Meng and 
Wessel~\cite{Meng} who studied the spin-1/2 easy-axis Heisenberg model on the 
SSL with ferromagnetic transverse spin exchange using quantum Monte-Carlo and 
degenerate perturbation theory. Besides the magnetization plateau at 1/3 of the 
saturated magnetization they found a further plateau at 1/2, which persists 
only in the quantum regime. The same results have been obtained by Liu and 
Sachdev analyzing the perturbative effects of the transverse fluctuations 
on the SSL spin multiplets with large easy-axis anisotropy~\cite{Liu}.

It should be noted that a similar behavior as for $TmB_4$ has been also 
observed for other rare-earth tetraborides. For example, for $ErB_4$ the
magnetization plateau has been found at $m/m_s=1/2$~\cite{Michi,Matas},
for $TbB_4$ at $m/m_s=1/2,4/9,1/3,2/9$ and $7/9$~\cite{Yoshii} and for
$HoB_4$ at $m/m_s=1/3,4/9$ and $3/5$~\cite{Matas}.   
%%%%%%%%%%%%%%%%%%
\section{Model}
%%%%%%%%%%%%%%%%%%
In the current paper we present an alternative model of stabilization the 
magnetization plateaus in the rare-earth tetraborides based on the fact 
that these materials, in contrast to $SrCu_2(BO_3)_2$, are metallic.
Thus for a correct description of ground-state properties of rare-earth 
tetraborides one should take into account both spin and electron subsystems
 as well as the coupling between them. Supposing that electron and spin 
subsystems interact only via the spin dependent Ising interaction $J_z$,
the Hamiltonian of the system can be written as
\begin{eqnarray}
H=\sum_{ij\sigma}t_{ij}d^+_{i\sigma}d_{j\sigma} 
+ J_z\sum_{i}(n_{i\uparrow}-n_{i\downarrow})S^z_i
\nonumber\\
- h\sum_i(n_{i\uparrow}-n_{i\downarrow}) + H_{JJ'}\ ,
\label{eq2}
\end{eqnarray}
where $d^+_{i\sigma}$, $d_{i\sigma}$  are the creation and annihilation 
operators of the itinerant electrons in the $d$-band Wannier state at site~$i$
and $n_{i\sigma}=d^+_{i\sigma}d_{i\sigma}$.
The first term of (2) is the kinetic energy corresponding to
quantum-mechanical hopping of the itinerant $d$ electrons between sites 
$i$ and $j$. These intersite hopping transitions are described by the 
matrix  elements $t_{ij}$, which are $-t$ if $i$ and $j$ are the nearest 
neighbors, $-t'$ if $i$ and $j$ are the next-nearest neighbors from the SSL  
and zero otherwise. The second term represents the above mentioned
anisotropic, spin-dependent local interaction of the Ising type between 
the localized spins and itinerant electrons. 
The third term describes an action of the magnetic field on the itinerant 
electrons.

To examine the magnetization curve corresponding to the model Hamiltonian
(2), we have used the well-controlled numerical method that we have elaborated 
recently to study the ground states of the spinless/spin-one-half
Falicov-Kimball model~
\cite{Farky1}. 
This method is described in detail in our previous
papers~\cite{Farky2,Cenci}
and thus we summarize here only the main steps of the algorithm:
(i) Chose a trial spin configuration $s=\{S^z_1,S^z_2, \dots ,S^z_L\}$.
(ii) Having $s$, $J_z$, $t$ and $t'$ fixed, find all eigenvalues
$\lambda^{\sigma}_k$ of $h_{\sigma}(s)=t_{ij}-\sigma J_zs_i\delta_{ij}$.
(iii) For a given $N=N_\uparrow+N_\downarrow$ (where $N$ is the total number
of electrons)
determine the ground-state energy
$E(s)=\sum_{\sigma}\sum_{k=1}^{N_\sigma}\lambda^{\sigma}_k
-h(N_\uparrow-N_\downarrow)+H_{JJ'}$
of a particular spin configuration $s$ by filling in the lowest
$N_\uparrow,N_\downarrow$ one-electron levels $\lambda^{\sigma}_k$.
(iv) Generate a new configuration $s'$ by flipping a randomly
chosen spin. (v) Calculate the ground-state energy $E(s')$.
If $E(s')<E(s)$ the new configuration is accepted, otherwise $s'$
is rejected. Then the steps (ii)-(v) are repeated until the convergence
(for given parameters of the model) is reached.
%%%%%%%%%%%%%%%%%%%
\section{Results and discussion}
%%%%%%%%%%%%%%%%%%%
%\\
Using the method discussed above we have performed exhaustive numerical
studies of the model (2) for a wide range of model parameters $h, J_z,
t, t'$ and $J/J'=1$ selected on the base of experimental
measurements~\cite{Gabani}. To exclude the finite size effects the numerical
calculations have been done for several different Shastry-Sutherland
clusters consisting of $L=8\times 8$, $10\times 10$ and $12\times 12$ sites.
The most important result obtained from these calculations is that the
ground-state spin arrangements exhibit the same structure for all examined
finite clusters. 
%%%%%%%%%fig2%%%%%%%%%%%%
%\vspace*{-0.5cm}
\begin{figure}[b]
\hspace*{-0.8cm}
\includegraphics[scale=0.5]{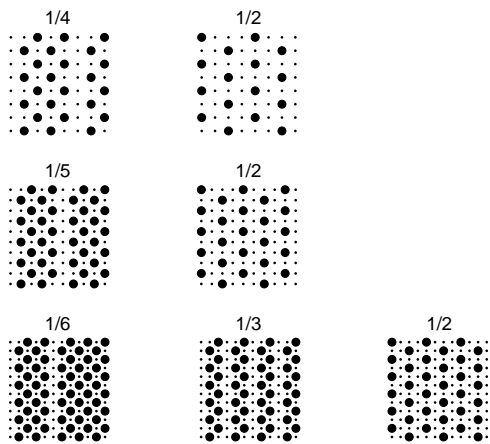}
\caption{ The complete list of the ground-state spin
configurations (for $0<m^{sp}/m_s^{sp}<1$) that are stable on finite 
intervals of $h$ for
$L=8\times8$, $L=10\times10$ and $L=12\times12$. The big (small) dots correspond
to the up (down) spin orientation.}
\end{figure}
%%%%%%%%%%%%
In general, this structure is formed by parallel
antiferromagnetic bands separated by ferromagnetic stripes and does not
depend on the anisotropic spin-dependent interaction $J_z$ as well as
nearest and next-nearest neighbor hopping integrals $t$ and $t'$. The
complete list of the ground-state spin arrangements (for $0<m^{sp}/m_s^{sp}<1$) 
that are stable on finite
intervals of magnetic field values are depicted on Fig.~2. 
The second
important observation is that the width $w$ of the antiferromagnetic bands
cannot be arbitrary, but fulfill severe restrictions. Indeed, we have found
that with exception the case $m^{sp}/m^{sp}_s=1/2$, in all remaining cases the
permitted
width of the antiferromagnetic band is only $w$ or $w+2$, where $w$ is the
even number. This fact is very important from the numerical point of view
since it allows us to perform the numerical calculations on much larger
clusters with the extrapolated set of configurations of the above described
type. 
The resulting magnetization curves obtained on the extrapolated set of
ground-state spin configurations consisting of parallel antiferromagnetic
bands of width $w$ ($w$ and $w+2$) separated by ferromagnetic stripes are
shown in Figs.~3-5  for selected values of model parameters, that represent the
typical behavior of the model. \\
%%%%%%%%%fig3%%%%%%%%%%%%
\vspace*{-0.5cm}
\begin{figure}[h]
\includegraphics[scale=0.4]{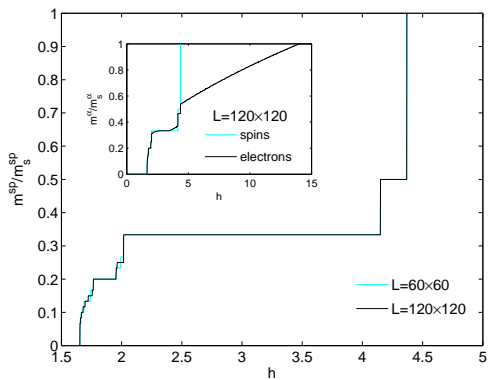}
\caption{
(Color online) Magnetization curves for $J'/J=1$, $J_z=2, t=4, t'=0$ 
and different values of $L$. 
Inset: magnetization curves of spin and 
electron subsystems ($\alpha=el$ or~$sp$).}
\end{figure}
%%%%%%%%%%%%%%%%%%%%%%%
%%%%%%fig4%%%%%%%%%%
\vspace*{-0.5cm}
\begin{figure}[ht]
\includegraphics[scale=0.4]{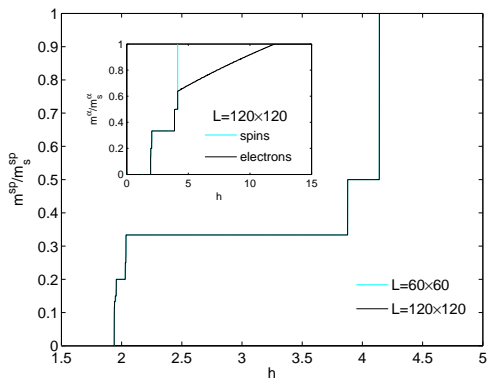}
\caption{
(Color online) Magnetization curves for $J'/J=1$, $J_z=4$, $t=4$, $t'=0$ 
and different values of $L$. 
Inset: magnetization curves of spin and 
electron subsystems.}
\end{figure}\\
%%%%%%%%%%%%%%
One can see that the switching on the
spin-dependent interaction $J_z$ between the electron and spin subsystems and
taking into account the electron hopping on the nearest ($t$) and
next-nearest ($t'$) lattice sites of the SSL leads to a stabilization of new
magnetization plateaus. In addition to the Ising magnetization plateau at
$m^{sp}/m^{sp}_s=1/3$ we have found two new magnetization plateaus
located at $m^{sp}/m^{sp}_s=1/2$  and $m^{sp}/m^{sp}_s=1/5$.
The ground-state spin arrangements
corresponding to these magnetization plateaus have the same structure
consisting of parallel antiferromagnetic bands of a width $w$ (where $w=1$
for $m^{sp}/m_s^{sp}=1/2$, $w=2$ for $m^{sp}/m_s^{sp}=1/3$ and  
$w=4$ for $m^{sp}/m_s^{sp}=1/5$) separated by ferromagnetic stripes.
Thus, our numerical results show
that besides the pure spin mechanism (e.g., the easy-axis Heisenberg
model on the SSL~\cite{Meng}) of stabilization the magnetization plateaus in
rare-earth tetraborides, there exists also an alternative mechanism based on
the coexistence of electron and spin subsystems that are present in these
materials. From this point of view it is interesting to compare in detail
the ground states obtained within these two different approaches. 
For $m^{sp}/m^{sp}_s=1/3$ our results are identical with ones
obtained within the Ising~\cite{Gabani, Chang} as well as 
easy-axis Heisenberg~\cite{Meng,Liu}
model on the SSL. The accordance between our and the easy-axis Heisenberg
solution~\cite{Meng} is found surprisingly also for $m^{sp}/m^{sp}_s=1/2$.
In this case both approaches predict the sequence of parallel
antiferromagnetic and ferromagnetic stripes. For  $m^{sp}/m^{sp}_s=1/5$
our results postulate a new type of spin ordering.\\
%%%%%%%%fig5%%%%%%%%%
\vspace*{-0.5cm}
\begin{figure}[h]
\includegraphics[scale=0.4]{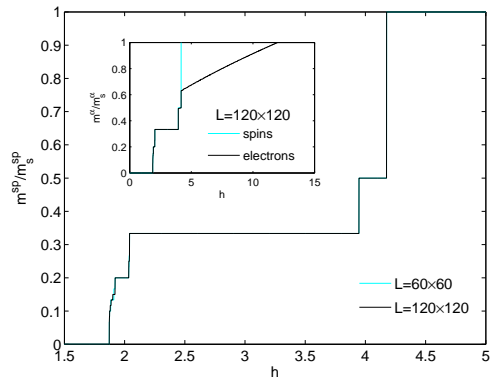}
\caption{
(Color online) Magnetization curves for $J'/J=1$, $J_z=4$, $t=4$, $t'=0.4t$ 
and different values of $L$. 
Inset: magnetization curves of spin and 
electron subsystems.}
\end{figure}
\\
%%%%%%%%%%%%%%%%
	While the magnetization plateaus at $m^{sp}/m^{sp}_s=1/2$ and 1/3
have been really found in the rare-earth
tetraborides~\cite{Gabani,Michi,Matas,Yoshii}, the
1/5-magnetization plateau in these compounds absent. Instead the 
1/5-magnetization plateau there have been observed magnetization plateaus at
smaller values of $m^{sp}/m^{sp}_s$, and namely, at $m^{sp}/m^{sp}_s=1/7$,
1/9 and 1/11 ($TmB_4$~\cite{Gabani}). Since the sizes of selected clusters ($60\times 60$ and
$120\times 120$) are not dividable by 7, 9 and  11 the absence of
magnetization plateaus at 1/7, 1/9 and 1/11 is nothing surprising. To verify
the possibilities of existence the magnetization plateaus at
$m^{sp}/m^{sp}_s=1/7, 1/9$ and 1/11 one has to examine much larger lattices.
Unfortunately, due to the numerical limitations we are able to study
clusters only slightly higher than 120 sites. Such cluster sizes (e.g.,
$L=140\times 140$) are sufficient for investigation the stability of
$m^{sp}/m^{sp}_s=1/7$ magnetization plateau, but they are too small for
verification the magnetization plateaus at $m^{sp}/m^{sp}_s=1/9$ and 1/11.
In Fig.~6 we present magnetization curves obtained on clusters consisting
of $70\times 70$ and $140\times 140$ sites together with the magnetization
curve for $L=120\times 120$. Comparing these results one can see that
a new magnetization plateau at $m^{sp}/m^{sp}_s=1/7$ is formed and that the
region of its stability is practically independent of $L$.\\ 
%%%%%%%fig6%%%%%%%%%%%
\vspace*{-0.3cm}
\begin{figure}[h]
\includegraphics[scale=0.4]{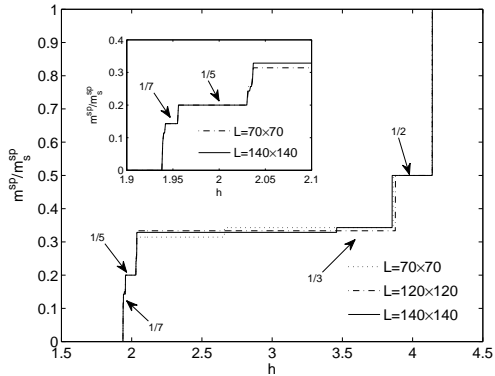}
\caption{
Magnetization curves for $J'/J=1$, $t=4$, $t'=0$, $J_z=4$ 
and different values of $L$. Inset: a detail of magnetization curves for
small $h$.}
\end{figure}\\
%%%%%%%%%%%%%%%%%%
Although we have
considered the clusters of different classes (the $70\times 70$ and
$140\times 140$ clusters are not dividable by 3 and the $120\times 120$ cluster
is not dividable by 7) the convergence of numerical results to 1/2, 1/3, 1/5
and 1/7 plateaus is apparent, what indicates that at least these plateaus
persist in the thermodynamic limit. It is not excluded that on much
larger clusters also the magnetic plateaus at 1/9 and 1/11 are stable. This
conjecture supports the fact that the 1/9 and 1/11 phases have the same type
of the ground-state spin ordering (consisting of parallel
antiferromagnetic bands (of the even width $w$) separated by ferromagnetic 
stripes) as the stable
phases corresponding to the magnetization plateaus at 1/3, 1/5 and 1/7.
However, with respect to our numerical results (presented above) showing on
the strong suppression of the plateau width with decreasing $m^{sp}/m^{sp}_s$ we
expect that the stability region of 1/9 and 1/11 phases will be very
narrow. 

	The magnetization process of the electron subsystem is very similar to one
described above for the spin subsystem, but only in the limit $m^{sp}/m^{sp}_s\leq
0.5$ (see Insets in Figs.~3-5). Indeed, we have found that for $m^{sp}/m^{sp}_s\leq 0.5$ the magnetization
curves of electron and spin subsystems fully coincide for the strong coupling
($J_z=4$) between electron and spin subsystems, and small deviations are
observed only for the intermediate coupling ($J_z=2$). However, a
different picture of magnetization processes of electron and spin
subsystems is observed in the limit $m^{sp}/m^{sp}_s>0.5$. In this limit the spin
subsystem is already fully saturated while the magnetization of the electron
subsystem changes continuously from $m^{el}/m_s^{el}=0.5$ to
$m^{el}/m_s^{el}=1$. 

	In summary, we have presented an alternative model of stabilization
the magnetization plateaus in rare-earth tetraborides based on the
coexistence of spin and electron subsystems (coupled by the anisotropic
spin-dependent interaction of the Ising type) in these materials. It was
shown that the switching on the spin-dependent interaction between the electron
and spin subsystems and taking into account the electron hopping on the
nearest and next-nearest lattice sites of the SSL leads to a stabilization of
magnetization plateaus at $m^{sp}/m_s^{sp}=1/2$, 1/3, 1/5 and 1/7 of the
saturated spin magnetization. The ground states corresponding to these
magnetization plateaus have the same structure consisting of parallel
antiferromagnetic bands of width $w=1$, 2, 4 and 6 separated by
ferromagnetic stripes. These results indicate that the electron subsystem
and its interaction with the spin subsystem can play the crucial role in the
correct description of magnetization processes in rare-earth tetraborides.
In our future work we plan to generalize this simple model by including the
long-range interactions (it was shown that such interactions suppress the
stability of the 1/3 phase~\cite{Suzuki}) and considering the Heisenberg spins
instead of the Ising ones. 
%%%%%%%%%%%%%%%%%%%%%%%%%%%%%%%%%%%%%%%%%%%%%%%%%%%%%%%%%%%%%%%%%%%%%%%%%%%
\section*{Acknowledgments} 
%%%%%%%%%%%%%%%%%%%%%%%%%%%%%%%%%%%%%%%%%%%%%%%%%%%%%%%%%%%%%%%%%%%%%%%%%%%
This work was supported by Slovak Grant Agency VEGA under Grant
No.2/0175/10, Slovak Research and Development Agency (APVV) under Grant 
VVCE-0058-07 and by the ERDF EU grant, under the contract No. ITMS26220120005.
H.C. acknowledges support of Stefan Schwartz
Foundation.
%%%%%%%%%%%%%%%%%%%%%%%%%%%%%%%%%%%%%%%%%%%%%%%%%%%%%%%%%%%%%%%%%%%%%%%%%%%%%
%                            REFERENCES                                     %
%%%%%%%%%%%%%%%%%%%%%%%%%%%%%%%%%%%%%%%%%%%%%%%%%%%%%%%%%%%%%%%%%%%%%%%%%%%%%
% Create the reference section using BibTeX
%\bibliography{nmr}
%%%%%%%%%%%%%%%%%%%%%%%%%%%%%%%%%%%%%%%%%%%%%%%%%%%%%%%%%%%%%%%%%%%%%%%%%%%%%

%

\end{document}